\documentclass[aps,prd,showpacs,showkeys,superscriptaddress,twocolumn]{revtex4-1}
\usepackage[colorlinks,linkcolor=blue,anchorcolor=blue,citecolor=blue,urlcolor=blue,breaklinks=true]{hyperref}
\usepackage{graphicx}
\usepackage{amsmath}
\usepackage{amssymb}
\usepackage{slashed}
\usepackage{latexsym}
\usepackage{epsfig}
\usepackage{amsbsy}
\usepackage{array}
\usepackage{amssymb}
\usepackage{setspace}
\usepackage{bm}
\usepackage{lipsum}
\usepackage{mathrsfs}
\usepackage{float}
\usepackage{color}
\usepackage{booktabs}
\usepackage[T1]{fontenc}
\usepackage{mathptmx}
\usepackage{changes}
\DeclareMathAlphabet{\mathcal}{OMS}{cmsy}{m}{n}
\DeclareSymbolFont{largesymbols}{OMX}{cmex}{m}{n}

\begin{document}
\author{Li-Qun Su}
\email{xzslq1203@smail.nju.edu.cn}
\affiliation{Department of physics, Nanjing University, Nanjing 210093, China}
\author{Chao Shi}
\email{cshi@nuaa.edu.cn}
\affiliation{Department of Nuclear Science and Technology, Nanjing University of Aeronautics and Astronautics, Nanjing 210016, China}
\author{Yong-Hui Xia}
\email{xiayh@xiahuhome.com}
\affiliation{Department of physics, Nanjing University, Nanjing 210093, China}
\author{Hongshi Zong}
\email{zonghs@nju.edu.cn}
\affiliation{Department of physics, Nanjing University, Nanjing 210093, China}
\affiliation{Department of physics, Anhui Normal University, Wuhu 241000, China}
\affiliation{Nanjing Proton Source Research and Design Center, Nanjing 210093, China}
\affiliation{Joint Center for Particle, Nuclear Physics and Cosmology, Nanjing 210093, China}
\date{\today}
	
\title{Color superconductivity with self-consistent NJL-type model}
	
\begin{abstract}
In this paper, the NJL-type model is used to investigate the color superconductivity. The four-fermion interactions of the NJL-type model are Fierz-transformed into two different classes, i.e., the quark-antiquark and the quark-quark interaction channels, associated with the chiral symmetry breaking and color superconductivity respectively. We conclude that the weighting factor between quark-antiquark and quark-quark interaction channels has significance on the phase structure when the mean-field approximation is employed, and the baryon number density gives a tight constraint on the weighting factor of quark-antiquark interaction channels. Besides, the susceptibilities show that the color superconducting phase transition is of the second-order and takes place before the chiral crossover transition as quark number density increases. In the end, we study the critical temperatures $T_c$ of the color superconductivity and it agrees with the perturbative result of diquark condensate $\Delta\approx0.57T_c$.
\bigskip

\end{abstract}
	
\maketitle
\section{Introduction}
It is believed that the strong interaction matter exhibits very rich phase structures at large baryon density, i.e., the hadronic matter is converted into the quark matter through a de-confinement phase transition and/or a phase transition of chiral symmetry restoration. It is known that, due to an arbitrary weak attraction, Fermi systems at low temperature will become Cooper instable \cite{BCS1957}. Such as electrons in opposite momentum and spins are paired below the critical temperature $T_c$, which lower the free energy of electrons. Inspired by the BCS theory, similar to the attraction between electrons caused by the phonons, the quarks can also be attractive to each other by gluons at low temperature and large baryon density. Thus, analogy to superconductivity, there exists color superconductivity at large density, which may be found in the center of compact stars \cite{Alford:2003kp,2001csqp.conf..137A,Ruester:2003zh,PhysRevD.92.105030,PhysRevC.71.045801,PhysRevD.69.045011}. Due to asymptotic freedom, the running coupling is weak enough and then the perturbative method can be applied so that quarks are in the BCS-type superconducting state \cite{Son1999,WilczekPRD1999,RischkePRD1999,RischkePRD2000}. In 1984, D. Bailin and A. Love employ the perturbative method to study color superconductivity, and the magnitude of the superconducting gap and critical temperature were found around 1 MeV \cite{BailinAndLove1983}. However, at the end of the 90s, the color superconductivity was investigated with effective field theories and the color superconducting gaps can reach the order of 100 MeV at quark chemical potential $\mu \sim 500 $ MeV \cite{AlfordAndWilczek1998,PhysRevLett.81.53,Buballa2005,RevModPhys.86.509,PhysRevD.72.054024,Buballa:2008zz}. Such huge gaps play a critical role in the structure of compact stars and the QCD phase diagram.

In this paper, the Nambu--Jona-Lasinio (NJL)-type model is used to discuss the color superconductivity. The interactions related to the color superconductivity can be obtained by the Fierz transformation to four-fermion interactions. In this way, the quark-quark interaction channels under mean-field approximation turn to diquark condensate, which can be regarded as the order parameter of color superconducting phase transition. In the previous works \cite{Buballa2005,RevModPhys.86.509,PhysRevD.72.054024,Buballa:2008zz}, the coupling constant of the quark-quark interaction channels is 3/4 times that of quark-antiquark interaction channels, i.e., $G_D = (3/4) G_s$ in \cite{PhysRevD.92.105030}, which can be obtained through Fierz transformation. This assumes the original quark-antiquark interaction channels and the Fierz transformed diquark interaction channels are equally combined. However, since the original Lagrangian and Fierz-transformed Lagrangian are mathematically identical, one can in principle take a linear combination of them with weighting factors $\alpha$ and $1-\alpha$ respectively. It should be noted here that the ``correct'' choice of the weighting factor $\alpha$ can be motivated only by physical reasoning, not by plain mathematics. Put differently, the Fierz transformation as such is exact, no matter what interaction channel we choose.
This is similar to the cases of \cite{wangqingwu,PhysRevD.100.094012,Wang2019,Li:2019ztm,Zhao:2019xqy}, in which the scalar interaction channel is Fierz-transformed to scalar and vector interaction channels.
As has been pointed out therein, this procedure brings change to the QCD phase transition property under the mean-field approximation, revealing the competition among different interaction channels within the mean-field approximation. Analogously, we employ the same method to investigate the color superconductivity by splitting the four-fermion interaction into two parts. One is transformed into quark-antiquark interaction channels, and the other into the form of quark-quark interaction channels. The proportion of the first part is set to be $\alpha$ and the later with $1-\alpha$.
Note that the quark-antiquark interaction channel is directly associated with the chiral property of QCD matter.
It gives rise to the chiral condensate, which is the order parameter of the chiral symmetry.  Meanwhile, the quark-quark interaction channel leads to the diquark condensate, which serves as the order parameter for superconductivity.   By varying the parameter $\alpha$, we study its influence on the phase transitions of QCD matter at high density and explore more possibilities.

This paper is organized as follows. 
In section \ref{section:2}, 
the effective Lagrangian of the NJL-type model is obtained, and the propagator of the effective Lagrangian with the parameter $\alpha$ is presented. 
In section \ref{section:3}, 
The chiral gap equation as well as the color superconducting gap equation with the parameter $\alpha$ are given. And the numerical results are shown in the diagrams.
In section \ref{section:4}, 
The effect of the temperature is investigated and the critical temperature of the color superconductivity is remarkable compared to the chemical potential.
In section \ref{section:5}, 
the conclusion is presented.

\section{Effective Lagrangian}\label{section:2}	
The Lagrangian of two flavor strong interacting matter from standard model is
\begin{align}
\mathcal{L}_\mathrm{QCD}=\bar{q}(i\gamma^{\mu}\partial_\mu -m)q+ g\bar{q}\gamma^\mu\lambda_aq A_\mu^a-\dfrac{1}{4}F^a_{\mu\nu}F^{\mu\nu}_a,\label{eq:1}
\end{align}
where $F_{\mu\nu}^{a}=\partial_\mu A_\nu^a-\partial_\nu A_\mu^a+gf^{abc}A_\mu^b A_\nu^c$ represents the gluon field strength tensor, $q$ represents the quark fields and $m$ is the current quark mass matrix.
Under the path integral, the gluon fields $A_\mu^a$ can be integrated out, and the effective Lagrangian with only one-gluon exchange four-fermion interaction are obtained \cite{Buballa2005}: 
\begin{align}
\mathcal{L}=\bar{q}(i\gamma^\mu\partial_\mu-m)q - g(\bar{q}\gamma^\mu\lambda_a q)^2.\label{eq:2}
\end{align}
In fact the four-fermion interaction terms contain all possible interaction channels. One can always employ the Fierz transformation to reveal these underlying interactions. For color superconductivity, the interactions of the effective Lagrangian is separated and Fierz-transformed into two ways, $\mathscr{F}_{\bar{q}q}=(\bar{q}\hat{O} q)^2~$and $\mathscr{F}_{qq}=(q\hat{O} q)^2$, related to the chiral phase transition and color superconductivity respectively. Although $\mathscr{F}_{\bar{q}q}\{\mathcal{L}\}$ and $\mathscr{F}_{qq}\{\mathcal{L}\}$ are mathematically identical, the diquark condensate via the mean-field approximation are evidently influenced by the ratio between two ways of Fierz transformation. This is due to the fact that Fierz transformation and mean-field approximation are not commutative.  
In order to evaluate the contributions from different interaction channels, the proportion $\alpha$ of original Lagrangian transforms into the quark-antiquark interaction channels, and the rest of original Lagrangian then  turns to quark-quark interaction channels which is multiplied by $(1-\alpha)$. Thus, the effective Lagrangian now becomes \cite{Buballa2005},
\begin{align}
\mathcal{L}=&\bar{q}(i\gamma^\mu\partial_\mu-m)q +\alpha\mathscr{F}_{\bar{q}q}[- g(\bar{q}\gamma^\mu\lambda_a q)^2]\nonumber\\
&+(1-\alpha)\mathscr{F}_{qq}[- g(\bar{q}\gamma^\mu\lambda_a q)^2].\label{eq:3}
\end{align}
In principle, $\alpha$ should be constrained by experiments rather than self-consistent mean-field approximation itself. But, due to the lack of relevant experimental data of strongly interacting matter, the real weighting factor $\alpha$ is uncertain, so it is set as a free parameter from zero to one in the present manuscript. Similarly to color superconductivity, the diquark condensate have to satisfy the Pauli principle. We only keep such terms where the operator between two fermion fields is asymmetric, where the operator in color space is in color $\bar{3}$ channel \cite{BailinAndLove1983}.
\begin{align}
\mathcal{L}=&\bar{q}(i\gamma^\mu\partial_\mu-m+\mu\gamma^0)q \nonumber\\
&+\alpha\dfrac{N_c^2-1}{N_c^2}g\big[(\bar{q} q)^2-\dfrac{1}{2}(\bar{q}\gamma^0 q)^2\big]\nonumber\\
&+(1-\alpha)\dfrac{N_c+1}{2N_c}g(\bar{q}i\gamma_5 \tau_A\lambda_{A^{'}} q_c)(\bar{q}_c   i\gamma_5\tau_A\lambda_{A^{'}} q),\label{eq:4}
\end{align}
here, charge conjugations are introduced
\begin{align}
&q_c(x)=C\bar{q}^T(x),\nonumber\\&\bar{q}^T_c(x)=q^T(x)C.\label{eq:5}
\end{align}
We rewrite the effective Lagrangian:
\begin{align}
\mathcal{L}=&\dfrac{1}{2}\big[\bar{q}(i\gamma^\mu\partial_\mu-m+\mu\gamma^0)q + \bar{q}_c(-i\gamma^\mu\partial_\mu-m-\mu\gamma^0)q_c\big]\nonumber\\
&+\alpha\dfrac{N_c^2-1}{N_c^2}g\big[(\bar{q} q)^2-\dfrac{1}{2}(\bar{q}\gamma^0 q)^2\big]\nonumber\\
&+(1-\alpha)\dfrac{N_c+1}{2N_c}g(\bar{q}i\gamma_5 \tau_A\lambda_{A^{'}} q_c)(\bar{q}_c   i\gamma_5\tau_A\lambda_{A^{'}} q).\label{eq:6}
\end{align}
In order to obtain the thermodynamic properties of the quark matter, the mean-field approximation is employed:

\begin{align}
\mathcal{L}=&\dfrac{1}{2}\big[\bar{q}(i\gamma^\mu\partial_\mu-M+\tilde{\mu}\gamma^0)q + \bar{q}_c(-i\gamma^\mu\partial_\mu-M-\tilde{\mu}\gamma^0)q_c\big]\nonumber\\
&+\dfrac{1}{2}\big[\bar{q}_c(-\Delta^*)\gamma_5\tau_A\lambda_{A^{'}}q+\bar{q}\Delta\gamma_5\tau_A\lambda_{A^{'}}q_c\big]\nonumber\\
&-G_s\langle\bar{q}q\rangle^2+G_v\langle\bar{q}\gamma^0 q\rangle^2-H\langle\bar{q}i\gamma_5 \tau_A\lambda_{A^{'}} q_c\rangle\langle\bar{q}_c   i\gamma_5\tau_A\lambda_{A^{'}} q\rangle,\label{eq:7}
\end{align}
where $G_s=\alpha\dfrac{N_c^2-1}{N_c^2}g,~
G_v=\dfrac{1}{2}\alpha\dfrac{N_c^2-1}{N_c^2}g,~
H=(1-\alpha)\dfrac{N_c+1}{2N_c}g
$, and also
\begin{align}
&M=m-2G_s\langle\bar{q}q\rangle,\label{eq:8}\\
&\tilde{\mu}=\mu-2G_v\langle\bar{q}\gamma^0 q\rangle,\label{eq:9}\\
&\Delta^*=2H\langle\bar{q}\gamma_5 \tau_A\lambda_{A^{'}} q_c\rangle,\label{eq:10}\\
&\Delta=-2H\langle\bar{q}_c   \gamma_5\tau_A\lambda_{A^{'}} q\rangle.\label{eq:11}
\end{align}
It is clear that the parameter $\alpha$ determines the intensities of chiral condensate and diquark condensate. With the decrease of the $\alpha$, the color superconducting gap increases, and it means that the quark-quark interaction channels dominates the quark system.
We can define a bispinor field,
\begin{align}
\Psi(x)=\dfrac{1}{\sqrt{2}}\begin{pmatrix}
        q(x)\\
q_c(x)\\\end{pmatrix}\label{eq:12}
\end{align}
So, in the momentum space,
\begin{align}
\mathcal{L}=&\bar{\Psi}S^{-1}\Psi+V,\label{eq:13}
\end{align}
where V is the interaction potential,
\begin{align}
V=&-G_s\langle\bar{q}q\rangle^2+G_v\langle\bar{q}\gamma^0 q\rangle^2-H\langle\bar{q}i\gamma_5 \tau_A\lambda_{A^{'}} q_c\rangle\langle\bar{q}_c   i\gamma_5\tau_A\lambda_{A^{'}} q\rangle,\label{eq:14}
\end{align}
and the inverse of the propagator matrix is
\begin{align}
S^{-1}=
\begin{pmatrix}
\slashed{p}+\tilde{\mu}\gamma^0-M & \Delta\gamma_5\tau_A\lambda_{A'}\\
(-\Delta^*)\gamma_5\tau_A\lambda_{A'} & \slashed{p}-\tilde{\mu}\gamma^0-M\\
\end{pmatrix}.\label{eq:15}
\end{align}
Following the Pauli principle, the diquark condensate demands the operator between two fermion fields in the diquark condensate to be asymmetric in Dirac, flavor, and color space altogether. Hence we have diquark condensate $\Delta \propto \epsilon_{ij}\epsilon^{\alpha\beta 3}$, where the Latin indices represent the flavors and the Greek indices signify the colors. The number "3" indicates the choice of direction in color space. For simplicity, we choose the blue as the preferred direction, and the diquark condensate tells us that the color symmetry is broken from SU(3) to SU(2). Thus, the asymmetric operators are $\tau_A=\tau_2, ~\lambda_{A}=\lambda_{2}$. The propagator is 
\begin{align}
&S_{r}=\dfrac{(\slashed{p}^-_r+M_r)\big[(\slashed{p}^+_r+M_r)(\slashed{p}^-_r-M_r)-\Delta^2\big]}{\big[p_0^2-\omega_+^2\big]\big[p_0^2-\omega_-^2\big]}P_{r}^c\nonumber\\
&S_{b}=\dfrac{(\slashed{p}^+_b+M_b)}{\big[p_0^2-E_+^2\big]\big[p_0^2-E_-^2\big]}P_{b}^c, \label{eq:16}\nonumber\\
\end{align}
where $E_p^2=\boldsymbol{p}^2+M_r^2,~\omega_{\pm}^2=(E_p\pm\tilde{\mu})^2+\Delta^2=E_{\pm}^2+\Delta^2$, and $P_{r}~$and $P_{b}~$ are the projectors on the red/green and the blue sector in color space, respectively.
\section{Thermodynamic properties}\label{section:3}	
In the finite temperature field theory, the thermodynamic potential \cite{Sun:2007fc,Huang:2002zd,He:2006vr} of quark matter is given by the propagator,
\begin{align}
\Omega(T,\mu)=-T\sum_n\int\dfrac{d^3p}{(2\pi)^3}\dfrac{1}{2}\mathrm{Tr~ln}\big[\dfrac{1}{T}S^{-1}(i\omega_n,\boldsymbol{p})\big]-V.\label{eq:17}
\end{align}
The self-consistent solutions of these condensate correspond to the stationary points of the thermodynamic potential,
\begin{align}
\dfrac{\delta\Omega}{\delta\sigma}=\dfrac{\delta\Omega}{\delta\tilde{\mu}}=\dfrac{\delta\Omega}{\delta\Delta}=0.\label{eq:18}
\end{align}
We define the condensate $\sigma=\langle\bar{q}q\rangle$, $n=\langle\bar{q}q\rangle$ and $\delta=\langle\bar{q}_c   \gamma_5\tau_2\lambda_{2^{'}} q\rangle=-\langle\bar{q}\gamma_5 \tau_2\lambda_{2^{'}} q_c\rangle$. These lead to gap equations \cite{Buballa2005},
\begin{align}
&\sigma_r=-4\int\dfrac{d^3p}{(2\pi)^3}\big[\dfrac{M(E_p-\tilde{\mu})}{2E_p\omega_-}\mathrm{tanh}(\frac{\omega_-}{2T})\nonumber\\
&~~~~~~~~~+\dfrac{M(E_p+\tilde{\mu})}{2E_p\omega_+}\mathrm{tanh}(\frac{\omega_+}{2T})\big],\label{eq:19}\\
&\sigma_b=-4\int\dfrac{d^3p}{(2\pi)^3}\big[\dfrac{M}{2E_p}\mathrm{tanh}(\frac{E_-}{2T})+\dfrac{M}{2E_p}\mathrm{tanh}(\frac{E_+}{2T})\big],\label{eq:20}\\
&n_r=4\int\dfrac{d^3p}{(2\pi)^3}\big[\dfrac{(\tilde{\mu}-E_p)}{2\omega_-}\mathrm{tanh}(\frac{\omega_-}{2T})\nonumber\\
&~~~~~~~~~+\dfrac{(\tilde{\mu}+E_p)}{2\omega_+}\mathrm{tanh}(\frac{\omega_+}{2T})\big],\label{eq:21}\\
&n_b=4\int\dfrac{d^3p}{(2\pi)^3}\big[\dfrac{1}{2}\mathrm{tanh}(\frac{E_+}{2T})-\dfrac{1}{2}\mathrm{tanh}(\frac{E_-}{2T})\big],\label{eq:22}\\
&\delta=-8\int\dfrac{d^3p}{(2\pi)^3}\big[\dfrac{\Delta}{2\omega_-}\mathrm{tanh}(\frac{\omega_-}{2T})+\dfrac{\Delta}{2\omega_+}\mathrm{tanh}(\frac{\omega_+}{2T})\big],\label{eq:23}
\end{align}
where the first two equations are the chiral condensate of red/green and blue quarks, and the form of the chiral condensate of the blue quarks is the same as the general NJL-type models.
The next two equations express the particle number densities of red/green and blue quarks.
The last equation manifests the diquark condensate of the quark matter, which influences the chiral condensate and particle number densities.
The effective quark mass is determined by
\begin{align}
&M=m-2G_s(\sigma_r+\sigma_g+\sigma_b), \label{eq:24}\\
&\tilde{\mu}=\mu-2G_v(n_r+n_g+n_b),\label{eq:25}\\
&\Delta=-2H\delta,\label{eq:26}
\end{align}
where the current quark mass is set as $m=5.5 ~\mathrm{MeV}$. The parameters $G_s,~ G_v~$ and $H$ are
\begin{align}
&G_s=\alpha\dfrac{N_c^2-1}{N_c^2}g, ~G_v=\dfrac{1}{2}\alpha\dfrac{N_c^2-1}{N_c^2}g,\label{eq:27}\\
&H=(1-\alpha)\dfrac{N_c+1}{2N_c}g.\label{eq:28}
\end{align}
The bare quark mass m, the coupling constant g, and the cutoff $\Lambda$ are set to fit the pion mass, pion decay constant, and the quark condensate. Here, we choose the set of parameters from Ref. \cite{Asakawa:1989bq}, where $m= 5.5~ \mathrm{MeV}, g=5.074 \times 10^{-6}~\mathrm{MeV}^{-2}$,
and the three-momentum cutoff $\Lambda=631~$MeV for regularization of ultraviolet divergences.
\begin{figure}[t]
	\centering
	\includegraphics[width=1\linewidth]{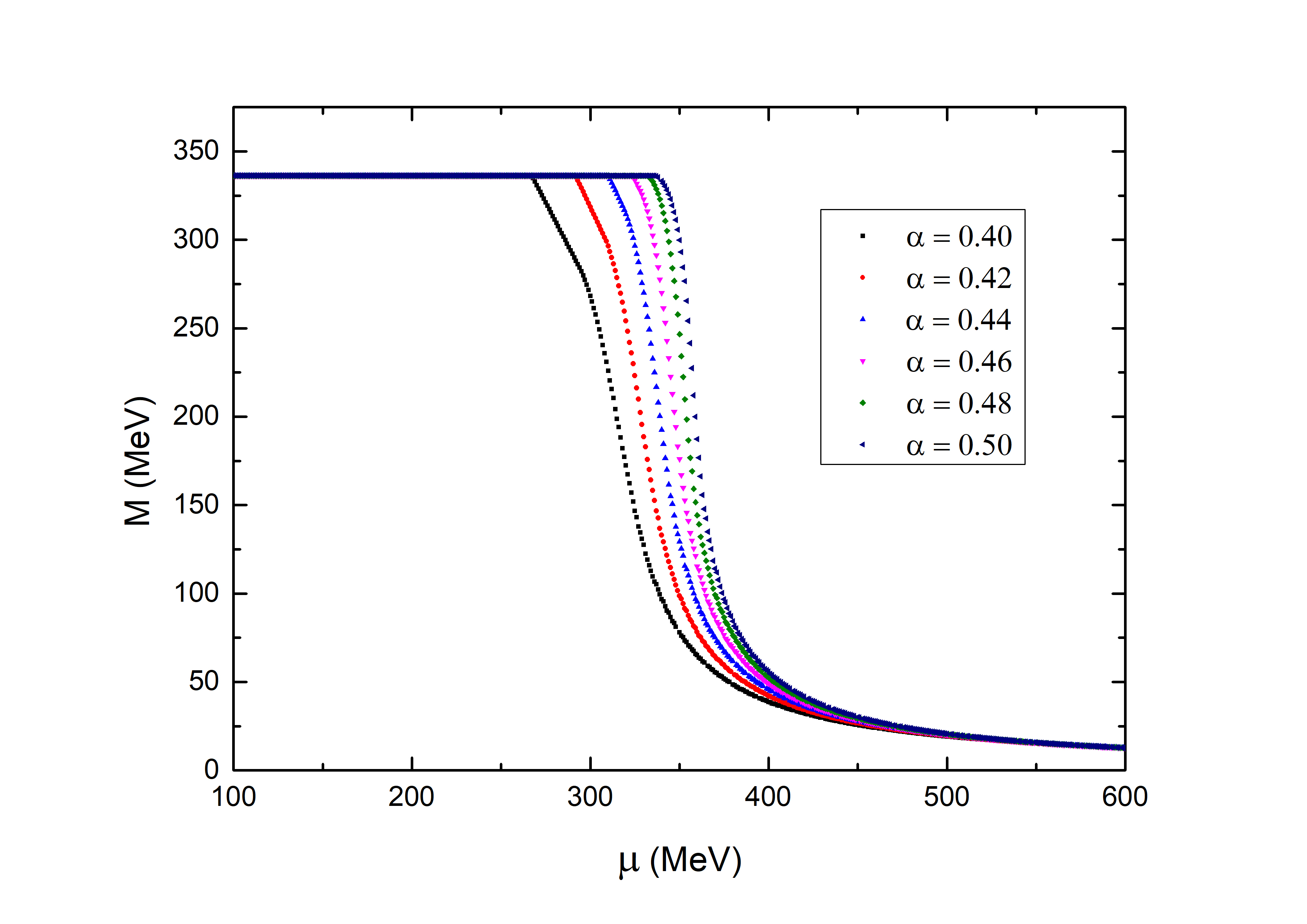}\\
	\caption{The effective quark mass as a function of the chemical potential at zero temperture. The effective quark mass start to fall at $\mu=268,~ 292,~ 311, ~324,~ 333, ~337~ \mathrm{MeV}$ with increase of $\alpha$ from 0.4 to 0.5. }\label{fig:mass}
\end{figure}

\begin{figure}[H]
	\centering
	\includegraphics[width=1\linewidth]{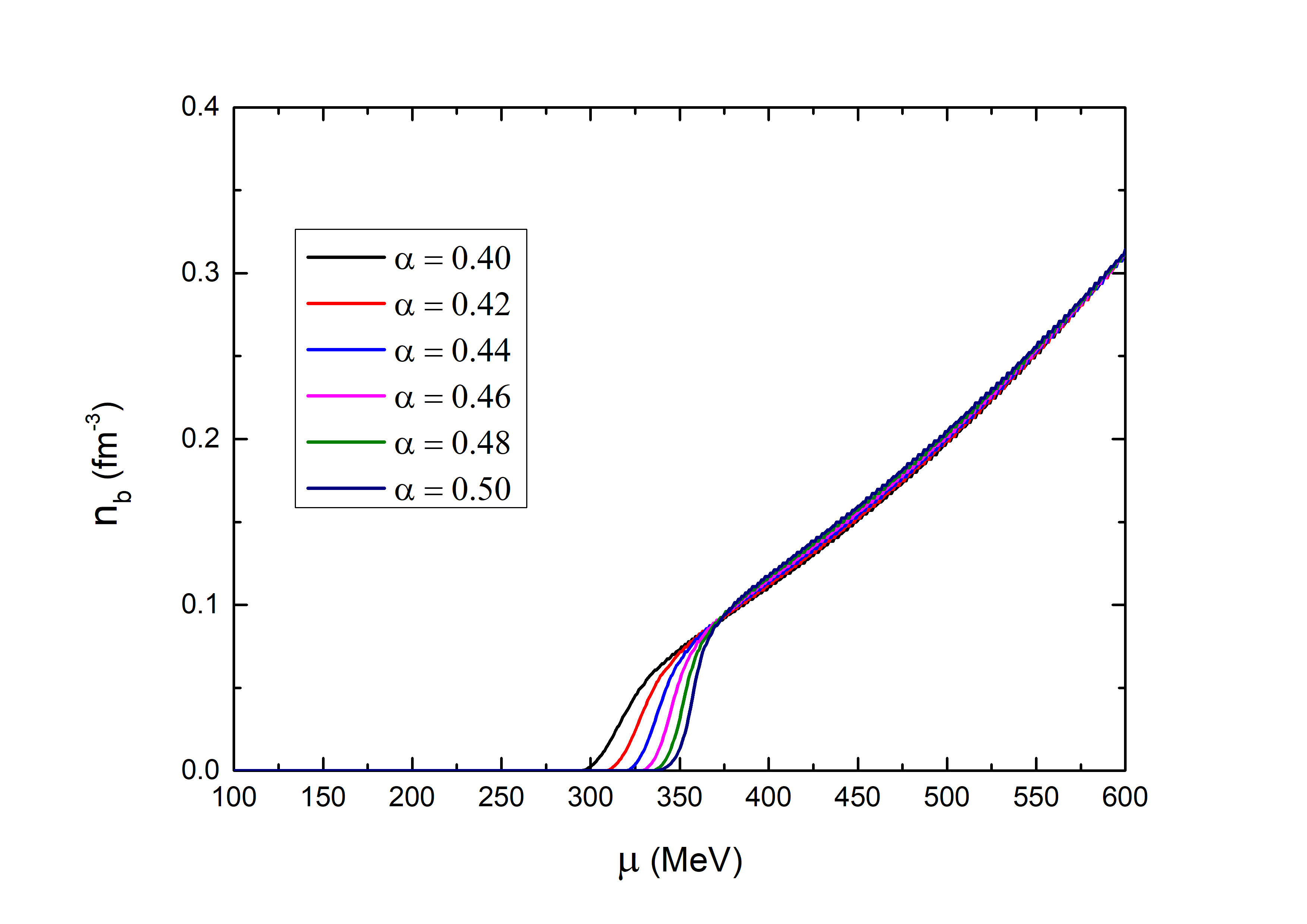}\\
	\caption{The particle number density of blue quarks as a function of the chemical potential.}\label{fig:nb}
\end{figure}

The effective quark mass with different $\alpha$'s at zero temperature and finite chemical potential are exhibited in Fig.~\ref{fig:mass}. Changing the $\alpha$ moves the curves. Here we let the $\alpha$ range from 0.4 to 0.5. When $\alpha$ is bigger than 0.5, the curves are not evidently influenced by the $\alpha$, and when the $\alpha$ is smaller than 0.4, the chiral transition starts to take place at $\mu < 300$ MeV, which is unphysical (See the discussion as the particle number density below). For these reasons, we set the range of $\alpha$ to be [0.4,~0.5].
\begin{figure}[]
	\centering
	\includegraphics[width=1\linewidth]{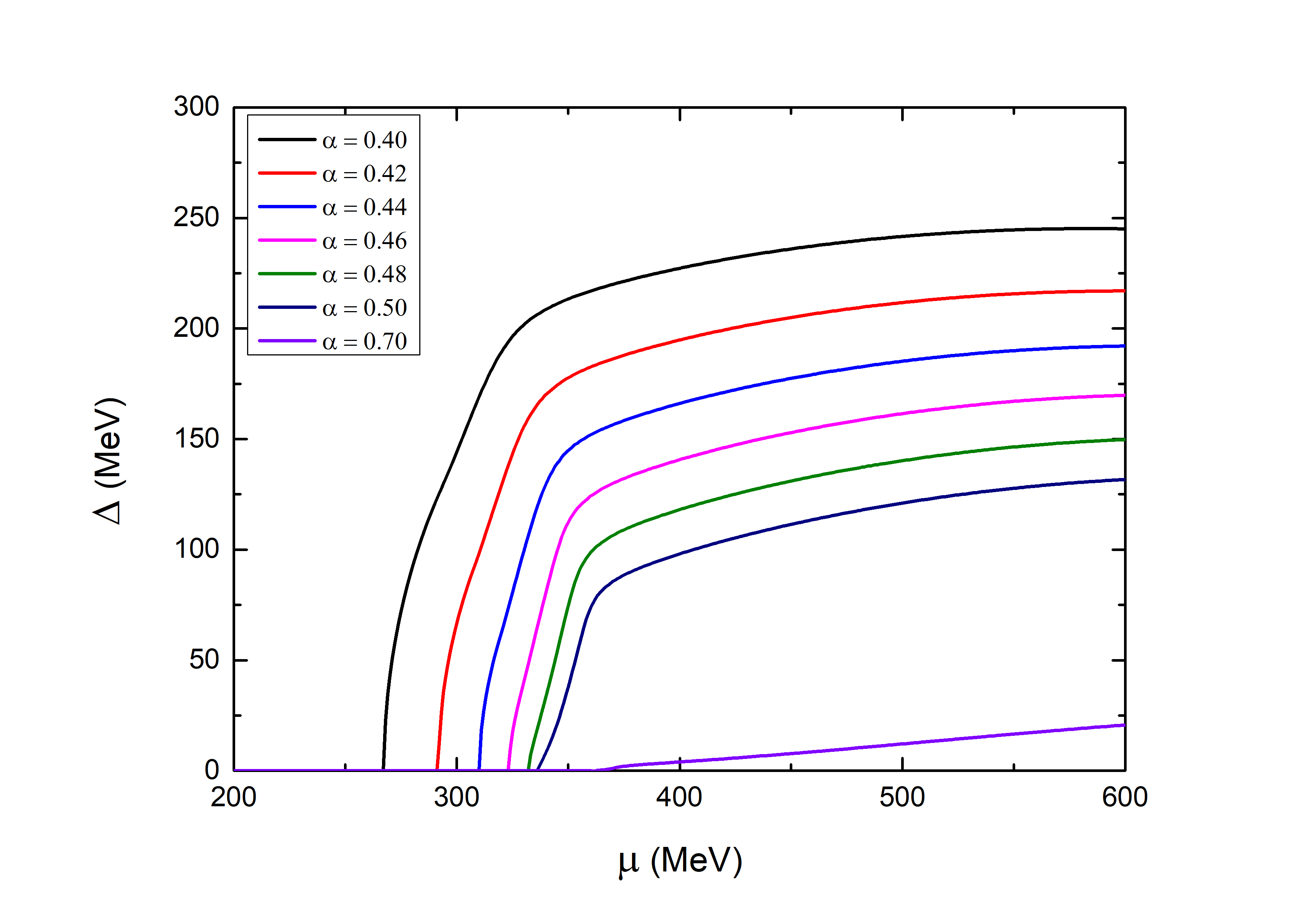}\\
	\caption{The color superconducting gaps as a function of the chemical potential at zero temperature are exhibited. The values of these gaps appear at $\mu=268, 292, 311, 324, 333, 337 ~\mathrm{MeV}$ with $\alpha$ ranging from 0.4 to 0.5. The maximum values of these gaps are $245, ~216,~219,~ 170, 150,~132~$MeV with different $\alpha$'s.}\label{fig:gap}
\end{figure}
As is shown in Fig.~\ref{fig:gap}, large intensities of diquark condensate not only enlarge the maximum value of the gap but also makes the diquark condensates show up at smaller a chemical potential. Physically, the particle number density should show up at $\mu \approx 313~$MeV, which is model-independent \cite{PhysRevD.58.096007}, where the nucleon is formed. But when the $\alpha$ is small enough, the color superconducting gaps show up at $\mu < 313$ MeV, hence unrealistic. Meanwhile, when $\alpha$ is too large, the diquark condensates are too small to observe. 
It is apparent that the stronger intensities of the diquark condensate make the declining parts of the effective quark mass take place at smaller chemical potentials.  Besides, the thermodynamic potentials are also reduced by the diquark condensate due to $\omega_{\pm}^2=E_{\pm}^2+\Delta^2$. For an analogy to the BCS theory, it means that the diquark condensate which has SU(2) color symmetry are related to a new state. Therefore, we take color superconducting gaps $\Delta$, corresponding to the diquark condensate, as the order parameter of the color superconducting phase transition. From Eq.~(\ref{eq:23}), the color superconducting gaps as the function of the chemical potential grow rapidly at the beginning and keep relatively steady with the increase of chemical potentials.
\begin{figure}[H]
	\centering
	\includegraphics[width=1\linewidth]{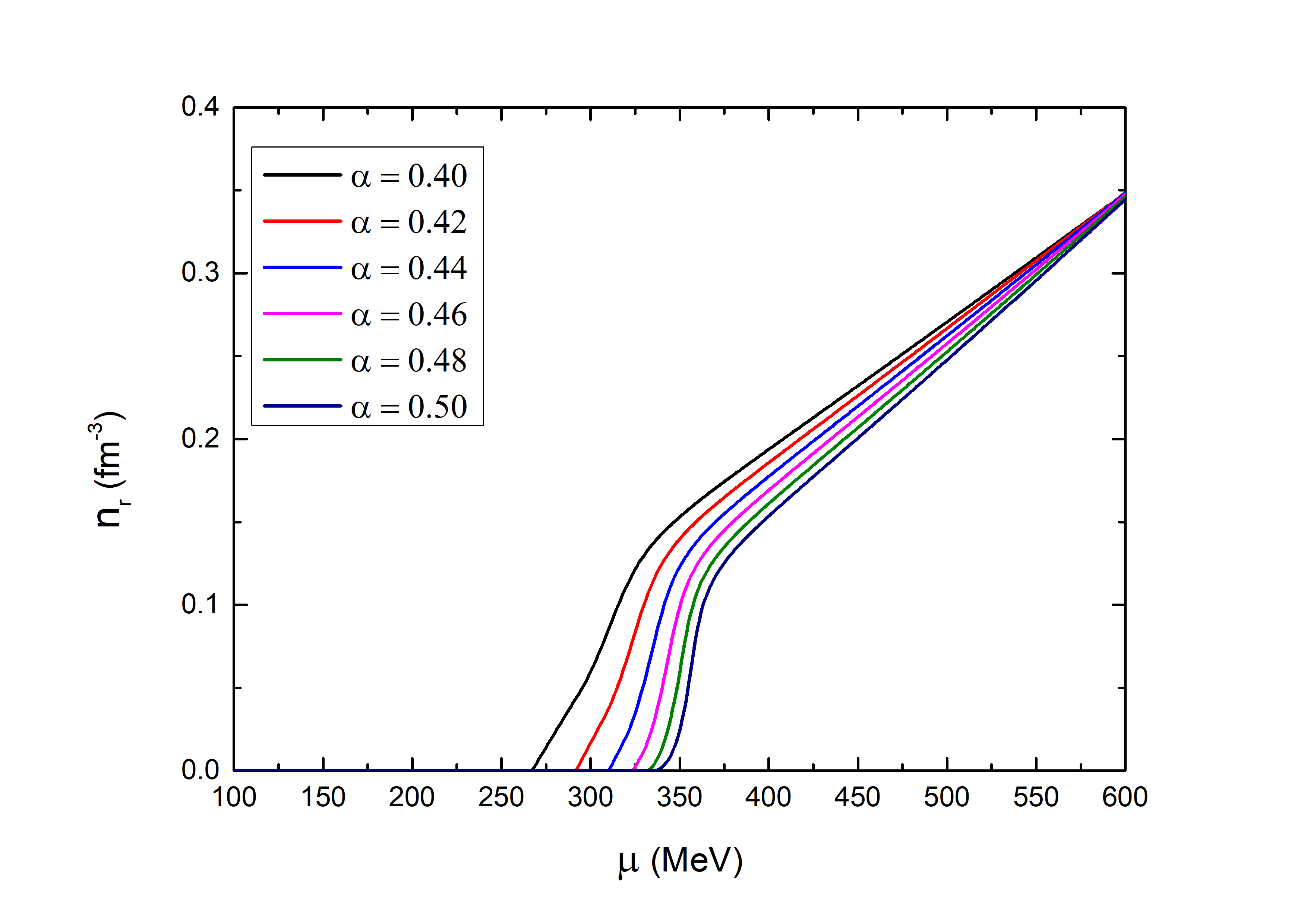}\\
	\caption{The particle number density of red/green quarks as a function of the chemical potential.}\label{fig:nr}
\end{figure}
The quark-quark interaction channels in the Lagrangian indicate the red/green quarks behave quite differently with the blue quarks, because of the dynamical breaking of color symmetry from SU(3) to SU(2). To see how quarks with different colors act, the particle number densities are essential. In Fig.~\ref{fig:nr} and Fig.~\ref{fig:nb}, the particle number density of the red/green quarks and blue quarks are presented respectively.
It shows that the smaller $\alpha$ which indicates larger diquark condensate lower the energy per quark, which makes the particles much easier to excite from the vacuum. As a result, the particle number densities will appear at very small chemical potentials. In addition, the influence of the diquark condensate on red/green quarks is much more evident than that on blue quark at medium chemical potential values. At higher chemical potentials, the influence of the $\alpha$ on quarks become insignificant for $\mu>600~$MeV. This is due to the use of the three momentum cutoff $\Lambda=631$ MeV in this paper, which specifies the scope of adaptation of the effective theory of this paper.
\begin{figure}[H]
	\centering
	\includegraphics[width=1\linewidth]{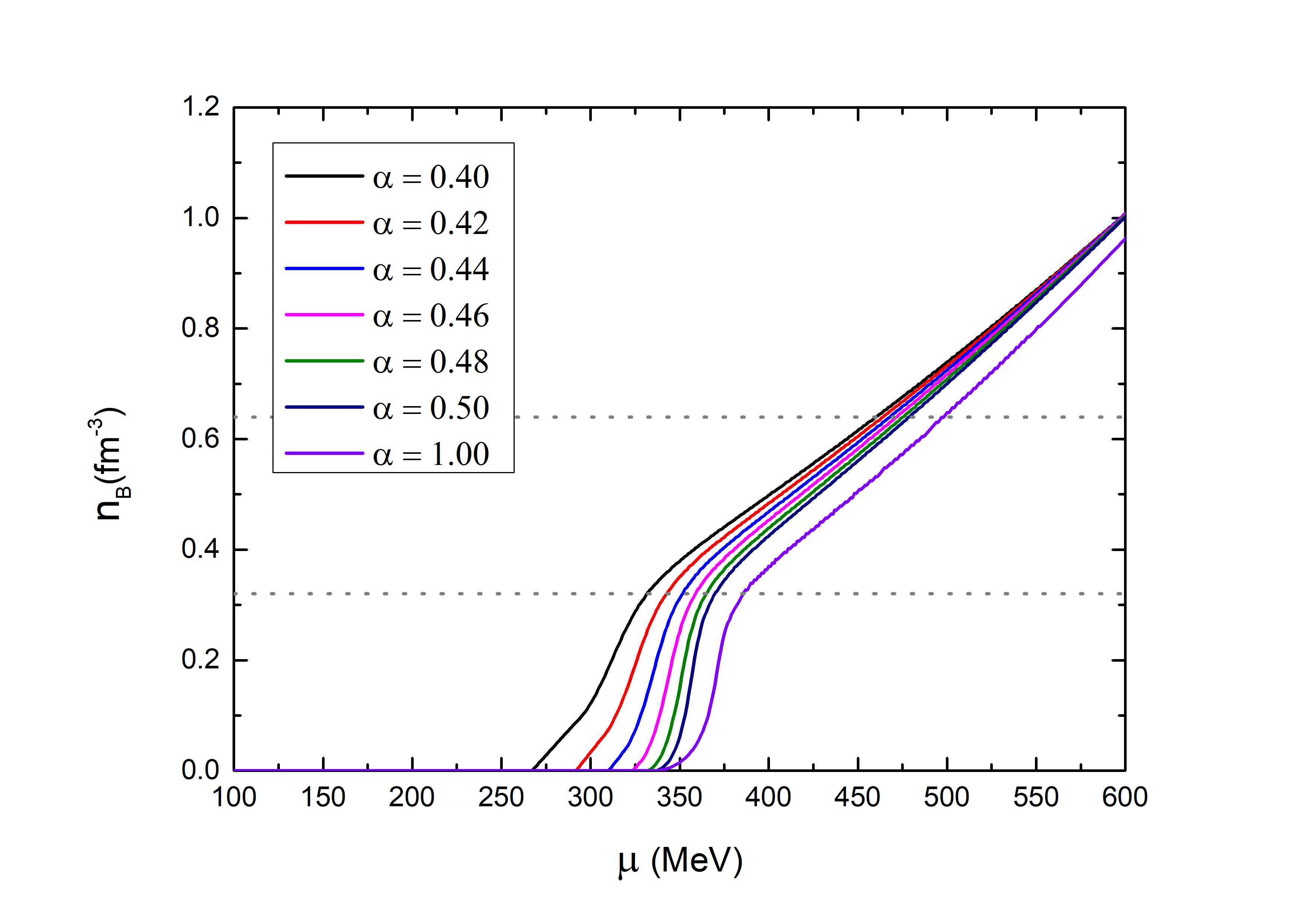}\\
	\caption{The baryon number density of quark matter as a function of the chemical potential. Here are two dashed lines which represent the region between $2n_0$ and $4n_0$}\label{fig:ntotal}
\end{figure}
The total densities of all three colors of quarks allow us to fit the experimental data of the nuclear saturation density, $n_0=0.16 ~\mathrm{fm}^{-3}$. The baryon number density of quark matter is defined:
\begin{align}
n_B=\dfrac {1}{3}(n_r+n_g+n_b),\label{eq:30}
\end{align}
where $n_B$ represents the baryon number density of quark matter. It is usually believed that the phase transition from hadrons to the deconfined quarks undergoes several regions \cite{Baym:2017whm}. For 
$n_B< 2n_0$, the dominating interactions are through a few exchanges of quarks and mesons, and the hadron degree of freedom is reliable at low densities. For $2n_0<n_B< (4\sim 7)n_0$, the many-quark exchanges occur and the hadron system is gradually percolated to the quark matter. For $n_B> (4\sim 7)n_0$, the description of the quark degree of freedom is valid and quarks are no longer confined in hadrons. 
The $2n_0$ and $4 n_0$ are plotted as the two horizontal dotted lines in Fig.~\ref{fig:ntotal}. Neglecting the diquark condensate which corresponds to the line of $\alpha=1$, the chemical potential ranges between $400~\mathrm{MeV}<\mu<500 ~\mathrm{MeV}$ for the region $2n_0<n_B< 4n_0$. After we take the diquark condensate into consideration, the baryon densities $n_B\sim2n_0$ is in the range between 300 MeV and 350 MeV.  Finally, for the most stable atomic nucleus, ${}^{56}\mathrm{Fe}$, the baryon number density should appear at $\mu\sim313~$MeV. Therefore, the parameter $\alpha$ is constrained to be over 0.44.

To study the order of chiral phase transition, we look into the chiral susceptibilities
\cite{PhysRevD.77.114028,Xu:2019ccc,Wang2019},
\begin{align}
\chi_r=\dfrac{\partial \sigma_r}{\partial m_r},\label{eq:31}\\
\chi_b=\dfrac{\partial \sigma_b}{\partial m_b},\label{eq:32}
\end{align}
where $m_r~$and$~ m_b~$represent the current quark mass of red/green and blue quarks respectively. 
\begin{figure}[t]
	\centering
	\includegraphics[width=1\linewidth]{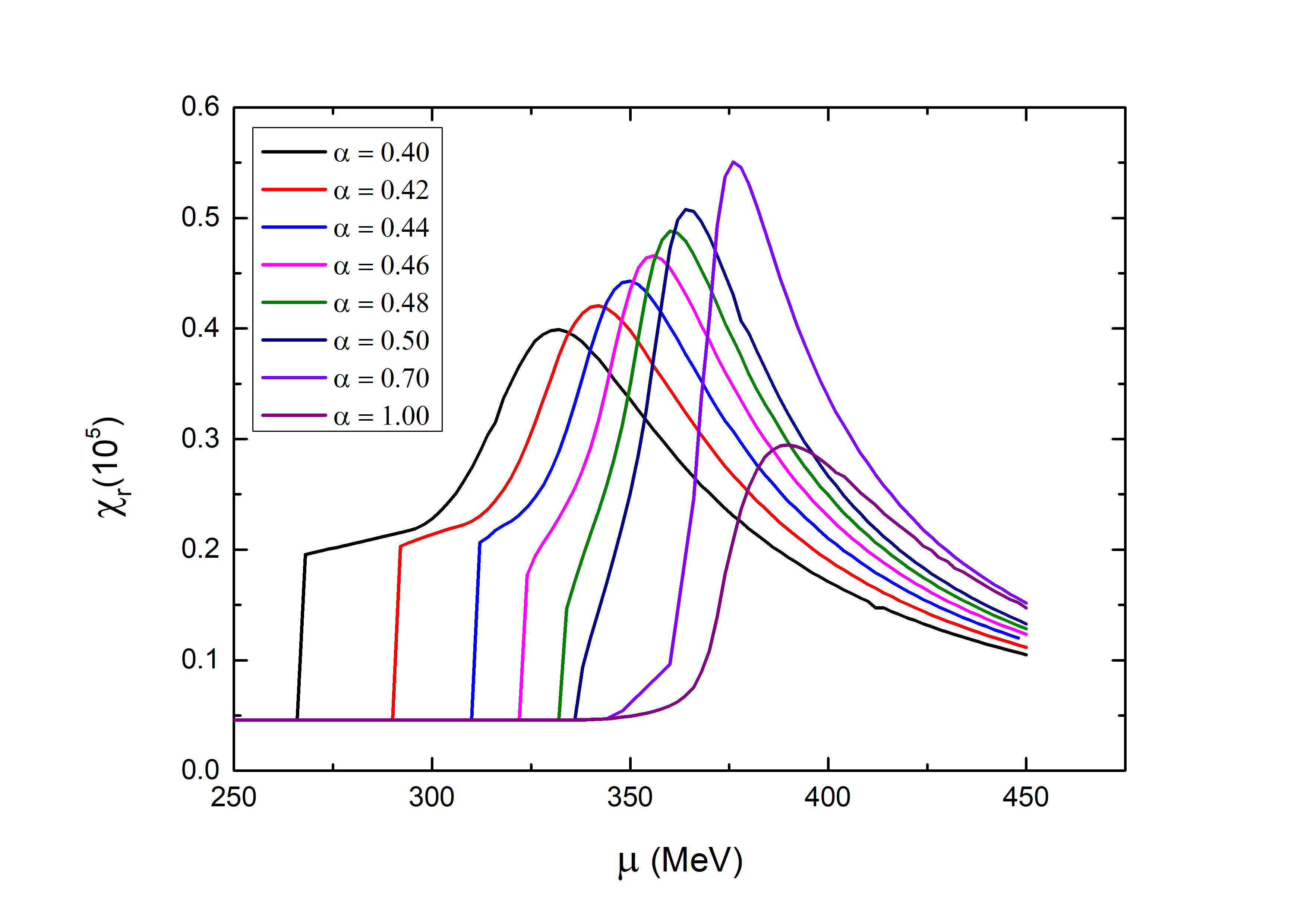}\\
	\caption{The chiral susceptibility of chiral condensate of red/green quarks.}\label{fig:susR}
\end{figure}
\begin{figure}[t]
	\centering
\includegraphics[width=1\linewidth]{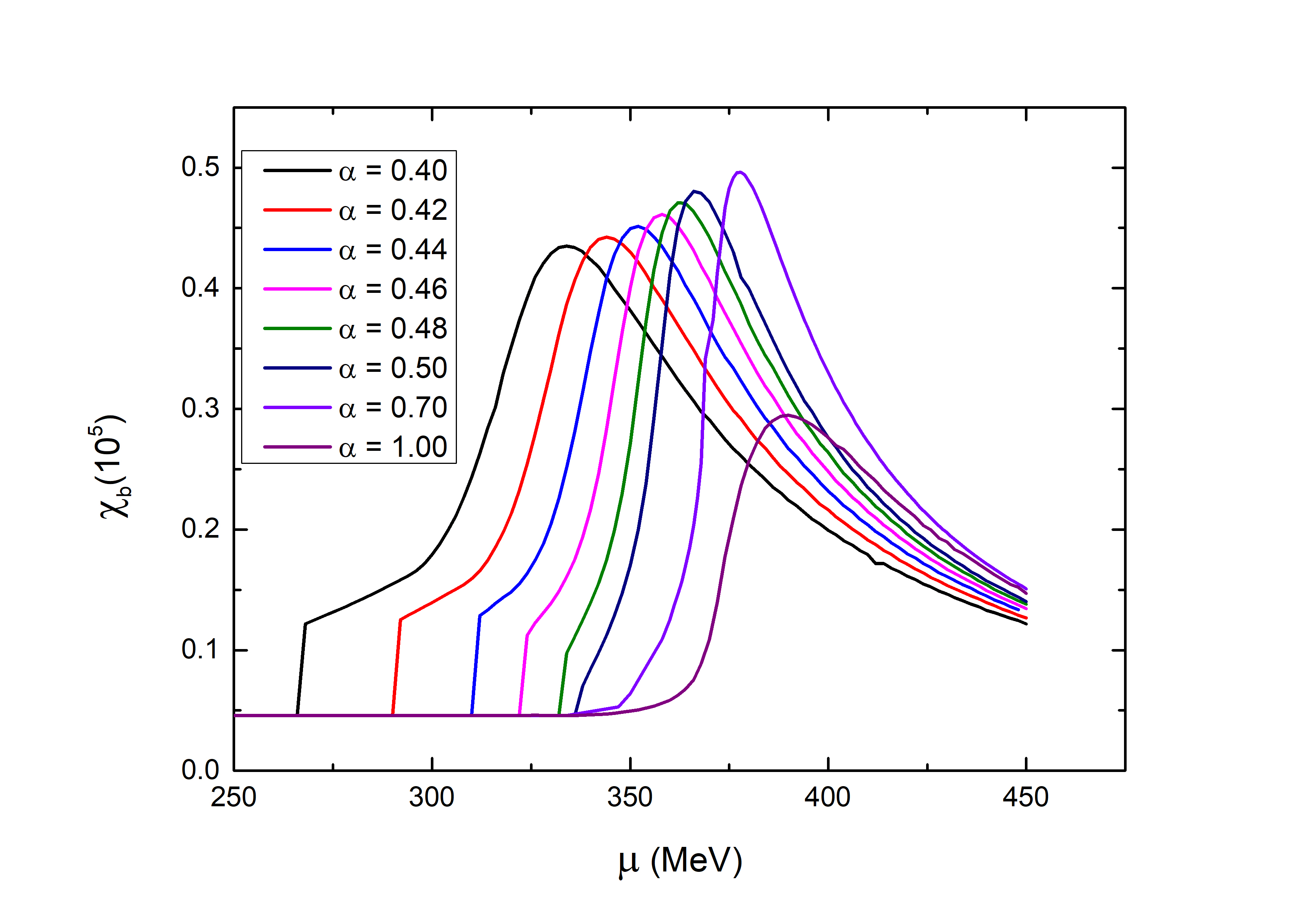}\\
	\caption{The chiral susceptibility of chiral condensate of blue quarks.}\label{fig:susB}
\end{figure}
The chiral susceptibility provides many useful information, as shown in Fig.~{\ref{fig:susR}} and Fig.~{\ref{fig:susB}}. First, the curves of chiral susceptibility show that the chiral transition takes the form of a crossover. Although in some studies \cite{Buballa2005,Wu:2017eon,Fan:2017wmb} the transition is a first order phase transition, it is found that when the vector interaction is strong enough, the first order phase transition turns to the crossover \cite{Buballa2005,Wang2019}. Second, the position of peak shifts toward smaller chemical potential as the parameter $\alpha$ decreases. Meanwhile, the maximum value of peaks of chiral susceptibilities of red or green quarks are prominently influenced by $\alpha$. This is because the gap equations Eq.~(\ref{eq:19}) and Eq.~(\ref{eq:21}) show that red or green quarks are directly affected by diquark condensates $\Delta$, while that of blue quarks from Eq.~(\ref{eq:20}) and Eq.~(\ref{eq:22}) are only indirectly connected.
 Interestingly, by comparing Fig.~{\ref{fig:ntotal}} and Fig.~{\ref{fig:susR}}, one finds the  chiral crossover transition point lies near the area when baryon number densities reach $2n_0$, where hadron system starts to percolate to the quark matter. Finally, we remark that for smaller $\alpha's$, there exists a plateaus before the peak of the chiral susceptibility in Fig.~{\ref{fig:susR}} and Fig.~{\ref{fig:susB}}, which is connected with an underlying phase transition of color symmetry breaking, as will be addressed below. 
\begin{figure}[H]
	\centering
	\includegraphics[width=1\linewidth]{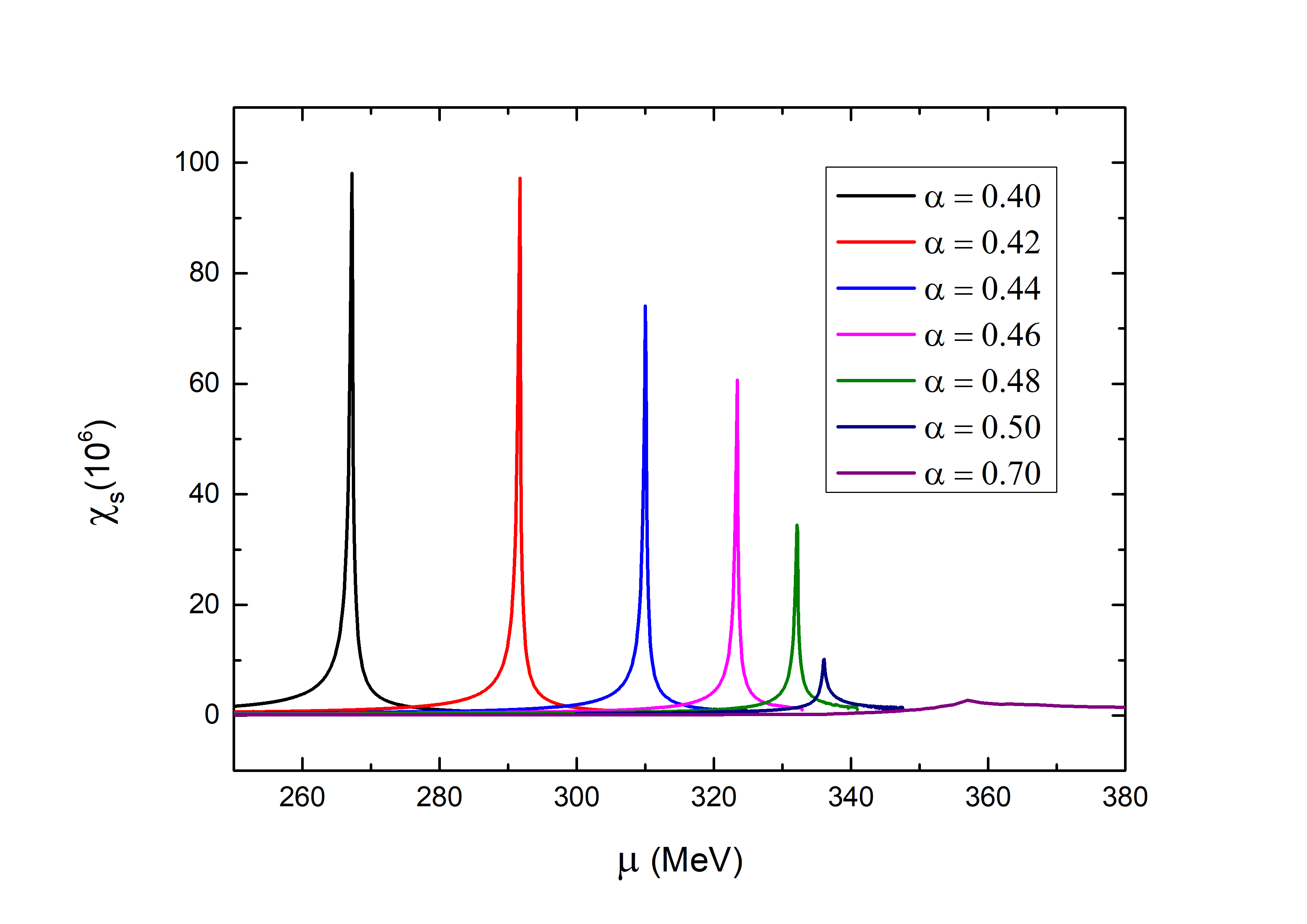}\\
	\caption{The color superconducting susceptibility.}\label{fig:susS2}
\end{figure}
To study the color superconducting phase transition, the interactions between the external field and quark matter are included, 
\begin{align}
\dfrac{1}{2}F\big[\bar{q}_c\gamma_5\tau_A\lambda_{A^{'}}q+\bar{q}\gamma_5\tau_A\lambda_{A^{'}}q_c\big],\label{eq:33}
\end{align}
where $F$ represents an external field related to the color superconductivity, which is the conjugate variable of diquark condensate. We define the susceptibility of color superconductivity:
\begin{align}
\chi_s=\lim_{F \rightarrow 0}\dfrac{\partial\langle\bar{q}_c \gamma_5\tau_A\lambda_{A^{'}} q\rangle}{\partial F}.\label{eq:34}
\end{align}
The susceptibilities of color superconductivity are shown in Fig.~{\ref{fig:susS2}}. These susceptibilities are numerically extremely sharp, and are actually diverging at the transition point, so the color superconducting phase transition is of the second-order. These peaks are at $\mu_s=267,~292,~310,~323,~332,~336~$MeV, which agrees with results in Refs. \cite{AlfordAndWilczek1998,PhysRevLett.81.53}, corresponding to the starting point of plateaus of chiral phase transition at $\mu_c=268,~292,~312,~324,~334,~338~$MeV in  Fig.~\ref{fig:susR} and Fig.~\ref{fig:susB} . Therefore, the plateaus of chiral susceptibilities indeed indicate the color superconducting phase transition. In conclusion, we find that the cold dense matter will  undergo color superconducting phase transitions first and then transit to the state with  chiral symmetry partially restored as the density increases.

\section{critical temperature}\label{section:4}
\begin{figure}[H]
	\centering
	\includegraphics[width=1\linewidth]{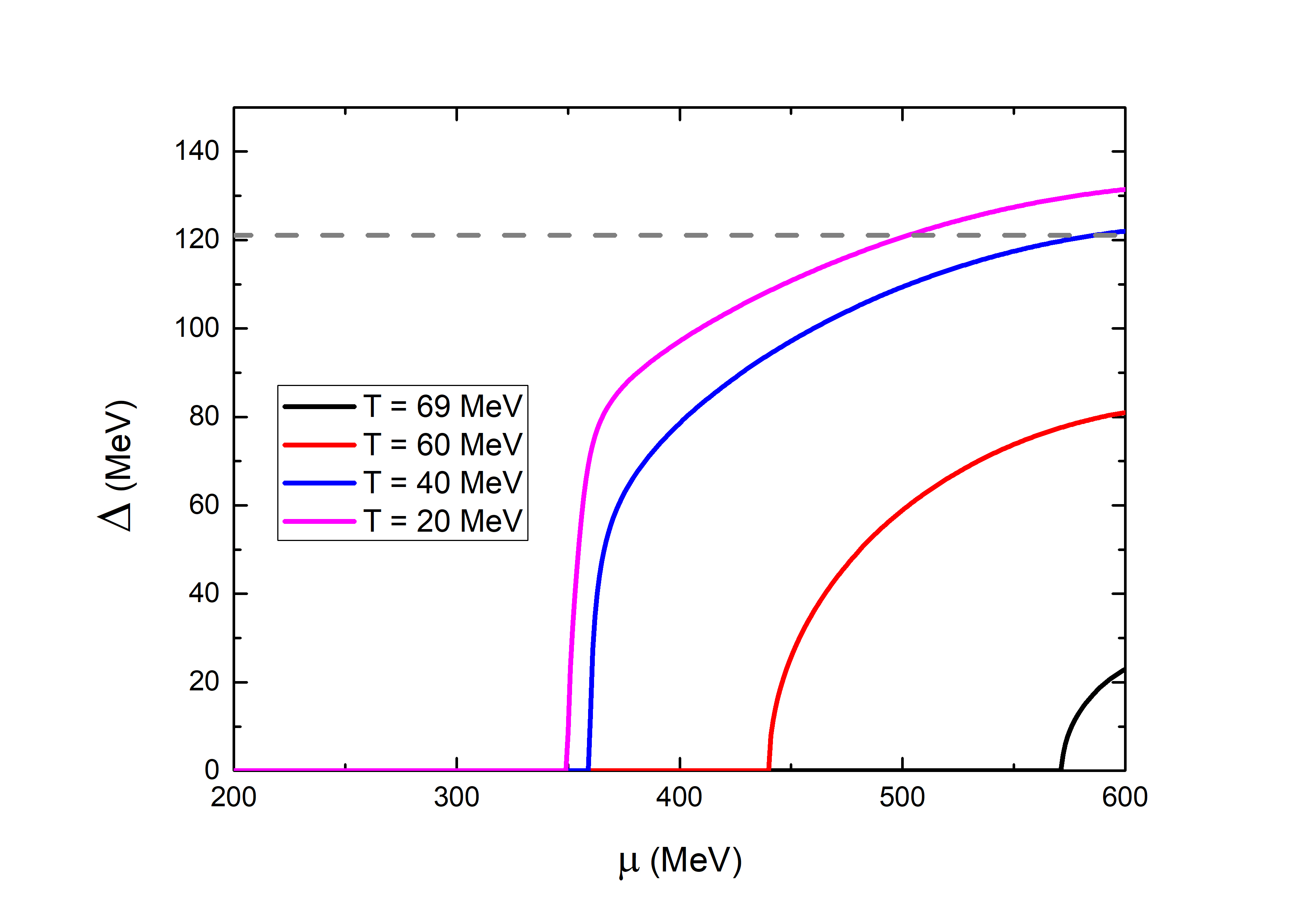}\\
	\caption{The color superconducting gaps as a function of the chemical potential with $\alpha = 0.50$, and the grey dash line represents the perturbative maximum color superconducting gap at zero temperature, corresponding to the critical temperature $\mathrm{T_c}=69 ~\mathrm{MeV}$.}\label{fig:T05}
\end{figure}

\begin{figure}[H]
	\centering
\includegraphics[width=1\linewidth]{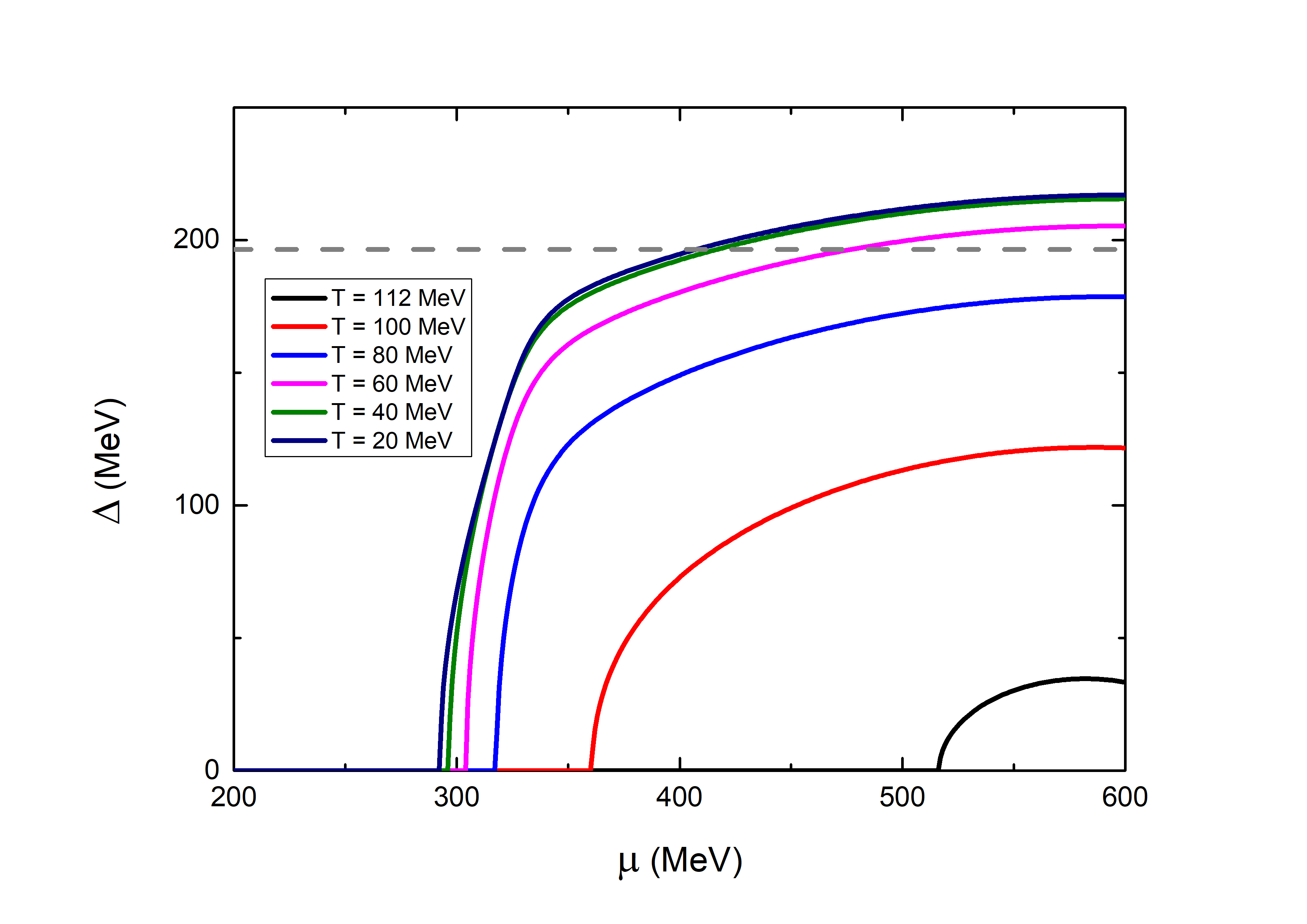}\\
	\caption{The color superconducting gaps as a function of the chemical potential with $\alpha = 0.42 $, and the grey dash line represents the perturbative maximum color superconducting gap at zero temperature, corresponding to the critical temperature $\mathrm{T_c}=112~\mathrm{MeV}$.}\label{fig:T042}
\end{figure}

We have shown that diquark condensate makes quark matter a more stable state at zero temperature. Next, we consider the case at finite temperature.
In BCS theory, the superconductivity is observed after the samples are cooled down to a certain critical temperature. One naturally wonders about the case in the color superconductor. Therefore we employ  Eq.~(\ref{eq:17}) at finite temperature.  
The color superconducting gaps at different temperatures are shown in Fig.~{\ref{fig:T05}} to Fig.~{\ref{fig:T042}} with three cases of the parameter $\alpha =$ 0.5, 0.46 and 0.42 respectively, which agree with the perturbative result $\Delta\approx0.57T_c$ \cite{PhysRevD.61.051501,PhysRevD.61.074017}, where $T_c$ is the critical temperature and $\Delta$ is the color superconducting gap at zero temperature (the grey dash lines ploted in Fig.~{\ref{fig:T05}}~to Fig.~{\ref{fig:T042}}). As the temperature goes up, the color superconducting gaps  decreases. The maximum temperature labeled in the diagrams are the critical temperatures of color superconductivity. When the temperatures exceed the critical point, the color superconducting gaps vanishes. 
\begin{figure}[H]
	\centering
\includegraphics[width=1\linewidth]{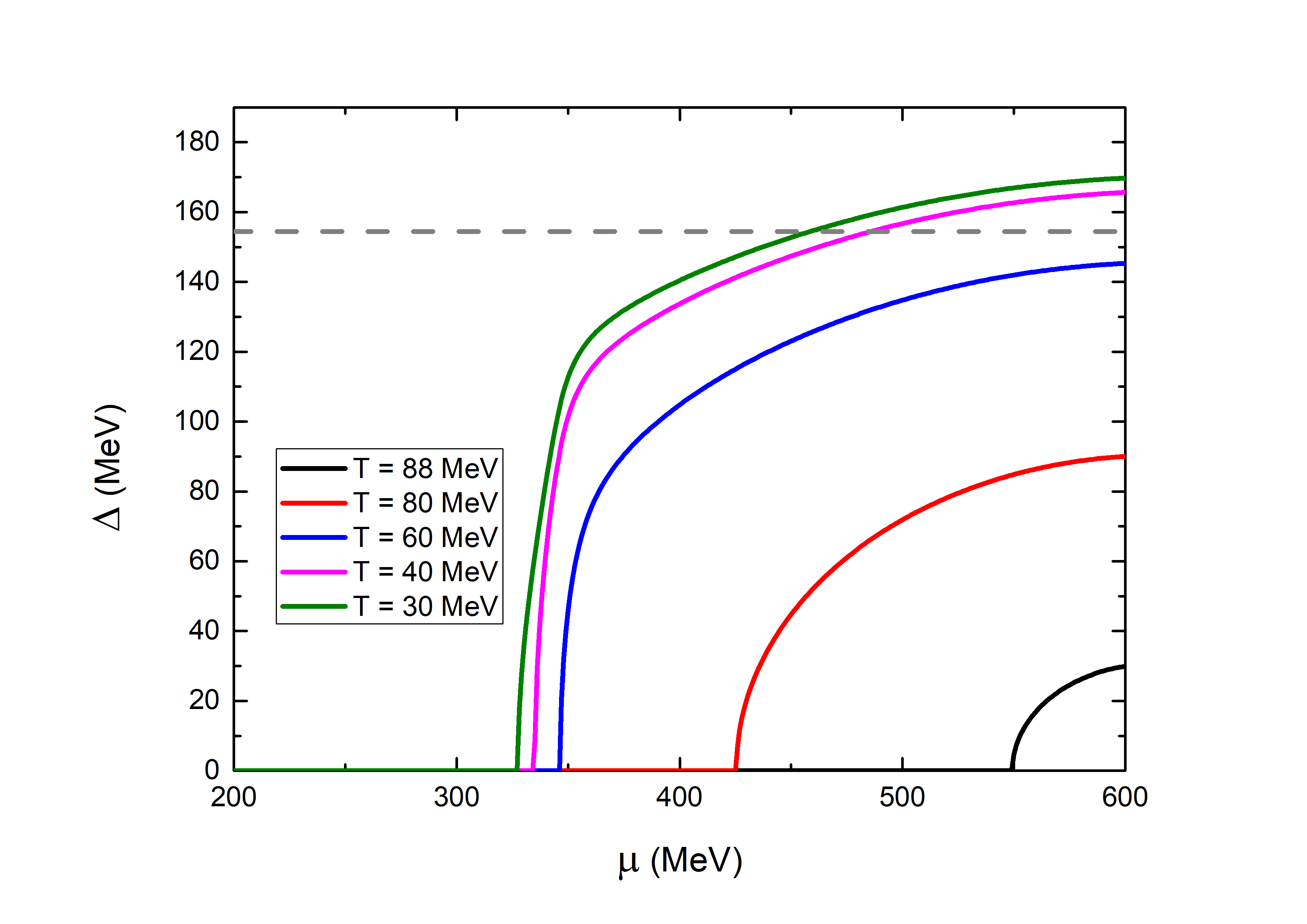}\\
	\caption{The color superconducting gaps as a function of the chemical potential with $\alpha = 0.46$, and the grey dash line represents the perturbative maximum color superconducting gap at zero temperature, corresponding to the critical temperature $\mathrm{T_c}=88~\mathrm{MeV}$.}\label{fig:T046}
\end{figure}

\section{summary and conclusion}\label{section:5}
In this paper, we discuss the chiral transition and color superconducting transition of cold dense matter at high density. The self-consistent NJL-type model is employed, and the Lagrangian is Fierz-transformed into two different channels, with a weighing factor $\alpha$ characterizing the partition. The chiral condensate and diquark condensate are obtained, corresponding to chiral transition and color superconducting transition. As an analogy to the superconductivity, the color superconducting gap decreases thermodynamic potential and leads to the color superconducting state. By introducing an external field, the susceptibility of color superconducting is employed. The superconducting transition turns out to be of second-order, and we find the chiral phase transition happens after color superconducting transition as the system gets denser. In the end, we study the color superconductor at finite temperature and find the critical temperatures of the color superconductor may reach around 88 MeV (corresponding to $\alpha=0.46$). 

Finally, we remark on the role of parameter $\alpha$. The $\alpha$ is set to measure quark-quark interaction channels and quark-antiquark interaction channels. With the decrease of the $\alpha$, diquark condensate associated with quark-quark interaction channels are stronger, and therefore easier to form color superconductivity. The original Lagrangian is not influenced by $\alpha$, but the mean-field approximation brings a difference. Hence, $\alpha$ can only be determined by experiments.  At present, we have shown that the $\alpha$ should be no less than 0.44, which leads to nonvanishing baryon number density that emerges at $\mu_c\approx 313$ MeV. More accurate constraints require observations and evidence from condense QCD matter such as compact stars.
For example, the x-ray and pulsar observations provide the measurements of the radius and the mass of compact stars respectively. Besides, the tidal deformability is constrained from the gravitational wave observation as well. These astronomical observations restrict the equation of state of quark matter to be neither too stiff nor too soft, which in turn constrain the range of weighting factor $\alpha$. It should be noted that compact stars are expected to have electric and color charge neutrality. Thus these neutral conditions should be taken into consideration to satisfy the astronomical observations for further researches.
\section*{Acknowledgements}
This work is supported in part by the National Natural Science Foundation of China (under Grants No. 11475085, No. 11535005, No. 11905104, and No. 11690030) and by Nation Major State Basic Research and Development of China (2016YFE0129300).

\bibliography{ref}
\end{document}